# Imaging-Derived Coronary Fractional Flow Reserve: Advances in Physics-Based, Machine Learning, and Physics-Informed Methods

Tanxin Zhu[1], Emran Hossen[1], Chen Zhao[2], Jingfeng Jiang[3], Michele Esposito[4], Jiguang Sun[5], Weihua Zhou[1,6]

1 Department of Applied Computing, Michigan Technological University, Houghton, MI, USA

2 Department of Computer Science, Kennesaw State University, Marietta GA, USA

3 Department of Biomedical Engineering, Michigan Technological University, Houghton, MI, USA

4 Department of Cardiology, Medical University of South Carolina, Charleston, SC, USA

5 Department of Mathematical Sciences, Michigan Technological University, Houghton, MI, USA

6 Center for Biocomputing and Digital Health, Institute of Computing and Cyber-systems, and Health Research Institute, Michigan Technological University, Houghton, MI, USA

Corresponding author:

Weihua Zhou, PhD, Tel: +1 906-487-2666

E-mail address: whzhou@mtu.edu

Mailing address: 1400 Townsend Dr, Houghton, MI 49931

**Abstract**

*Purpose of Review* Imaging-derived fractional flow reserve (FFR) is rapidly evolving beyond conventional computational fluid dynamics (CFD)-based pipelines toward machine learning (ML), deep learning (DL), and physics-informed approaches that enable fast, wire-free, and scalable functional assessment of coronary artery stenosis. This review synthesizes recent advances in computed tomography (CT)- and angiography-based FFR measurement, with particular emphasis on emerging physics-informed neural networks and neural operators (PINNs and PINOs), as well as key considerations for their clinical translation.

*Recent Findings* ML/DL approaches have markedly improved automation and computational speed, enabling prediction of pressure and FFR from anatomical descriptors or angiographic contrast dynamics. However, their real-world performance and generalizability can remain variable and sensitive to domain shift, due to multi-center heterogeneity, interpretability challenges, and differences in acquisition protocols and image quality. Physics-informed learning introduces conservation structure and boundary-condition consistency into model training, improving generalizability and reducing dependence on dense supervision while maintaining rapid inference. Recent evaluation trends increasingly highlight deployment-oriented metrics, including calibration, uncertainty quantification, and quality-control gatekeeping, as essential for safe clinical use.

*Summary* The field is converging toward imaging-derived FFR methods that are faster, more automated, and more reliable. While ML/DL offers substantial efficiency gains, physics-informed frameworks such as PINNs and PINOs may provide a more robust balance between speed and physical consistency. Prospective multi-center validation and standardized evaluation will be critical to support broad and safe clinical adoption.

## 1. Introduction

### 1.1 Clinical background and motivation

Coronary artery disease (CAD) remains a leading cause of morbidity and mortality worldwide. A central clinical challenge is to distinguish anatomically visible stenosis from functionally significant lesions that truly cause ischemia, particularly in intermediate disease, where visual assessment is subjective and variable. Pressure-wire-based fractional flow reserve (FFR), measured under maximal hyperemia, is widely regarded as the invasive reference standard for lesion-specific physiological assessment and for guiding revascularization decisions [1, 2].

Despite its clinical value, routine pressure-wire-based FFR is not universally adopted in daily practice. Practical barriers include wire manipulation, the need for hyperemic agents, additional procedure time and cost, and patient- or operator-related constraints [3]. These limitations have motivated the development of wire-free, imaging-derived physiological assessment—aiming to provide actionable functional information with reduced procedural burden while preserving lesion-level decision support.

In this review, we focus on non-invasive imaging-derived approaches that infer coronary hemodynamics and functional significance from routinely acquired imaging data, with an emphasis on methods that can realistically integrate into clinical workflows and provide reliable outputs across heterogeneous image quality, acquisition protocols, and lesion morphologies.

### 1.2 Imaging-based alternatives to invasive FFR

Imaging-derived functional assessment leverages anatomical and/or contrast-flow information to estimate pressure drop and physiological significance along the coronary tree. Among available modalities, clinical translation and large-scale evidence have been most prominent for computed tomography coronary angiography (CCTA) and invasive coronary angiography (ICA), each with distinct strengths and constraints.

**CCTA-based functional assessment (CT-derived FFR):** CCTA provides a patient-specific 3D representation of the coronary anatomy and plaque burden, enabling physiologic assessment across the vascular tree. CT-derived FFR has demonstrated diagnostic value relative to invasive FFR and has been positioned as a noninvasive strategy to improve patient triage beyond anatomy alone [3-6]. However, real-world feasibility can be limited by image quality and reconstruction challenges (e.g., motion artifacts, noise, heavy calcification or stent blooming), as well as sensitivity to segmentation and centerline uncertainties. Consequently, practical adoption depends not only on diagnostic accuracy but also on computability, quality control, and robust handling of imperfect inputs [7, 8]. Cardiac CT artifact mechanisms and mitigation strategies (including blooming and other physics-driven artifacts) have been comprehensively discussed elsewhere [9, 10].

**Angiography-based functional assessment (ICA-derived physiology):** ICA is routinely performed for diagnosis and procedural planning in the catheterization laboratory. A major clinical motivation for ICA-derived physiology is to provide wire-free functional information within the cath-lab time budget, supporting immediate decision-making without advancing a pressure wire. Quantitative flow ratio (QFR) and related approaches have shown strong agreement with invasive FFR in validation studies, and randomized evidence supports QFR-guided strategies [11, 12]. In practice, angiography-derived methods are highly dependent on acquisition conditions, including adequate projection separation, minimal foreshortening/overlap, and stable contrast opacification; real-world studies and head-to-head analyses have highlighted non-computability drivers and practical requirements. These workflow constraints make explicit quality control and failure-mode management integral to real-world performance [13, 14].

### 1.3 A taxonomy of modern imaging-derived FFR methods

To synthesize a rapidly growing literature, we organize imaging-derived FFR methods into three methodological families that reflect different trade-offs among fidelity, speed, interpretability, and robustness:

(i) **Computational Fluid Dynamics (CFD) and physics-based modeling:** Physics-based approaches estimate pressure and flow by solving governing equations on image-derived coronary geometries. They offer mechanistic consistency and interpretability, but clinical feasibility is often constrained by high computational cost and sensitivity to physiological assumptions, most notably boundary conditions and modeling of downstream resistance [15-18].

(ii) **Machine learning and deep learning:** ML-based approaches fully or partially replace the traditional physics pipeline with data-driven inference, enabling near-instant computation and high automation. However, clinical adoption is hindered by dataset shift—where variations in scanners, protocols, and lesion morphology degrade generalization. This has increased attention to integrating calibration and uncertainty estimation into the deployment pipeline, ensuring AI-driven diagnostics remain reliable in heterogeneous clinical environments [19-22].

(iii) **Physics-informed learning (PINNs and neural operators):** Physics-informed learning aims to bridge CFD and ML by embedding physical structure (e.g., conservation laws and boundary consistency) into learning and inference [23, 24]. Beyond classical pointwise PINNs, operator learning (e.g., Fourier neural operators and physics-informed neural operators) has emerged as a promising direction for improving generalization across geometries and boundary configurations with faster inference [25, 26], an attractive property for scalable hemodynamic prediction and centerline-level pressure/FFR profiling.

This taxonomy is intentionally workflow-aware: in real-world deployment, the practical value of an imaging-derived FFR method depends not only on average diagnostic metrics but also on computability, quality control, failure modes, and safe fallback strategies when inputs are insufficient or confidence is low [8, 13, 14].

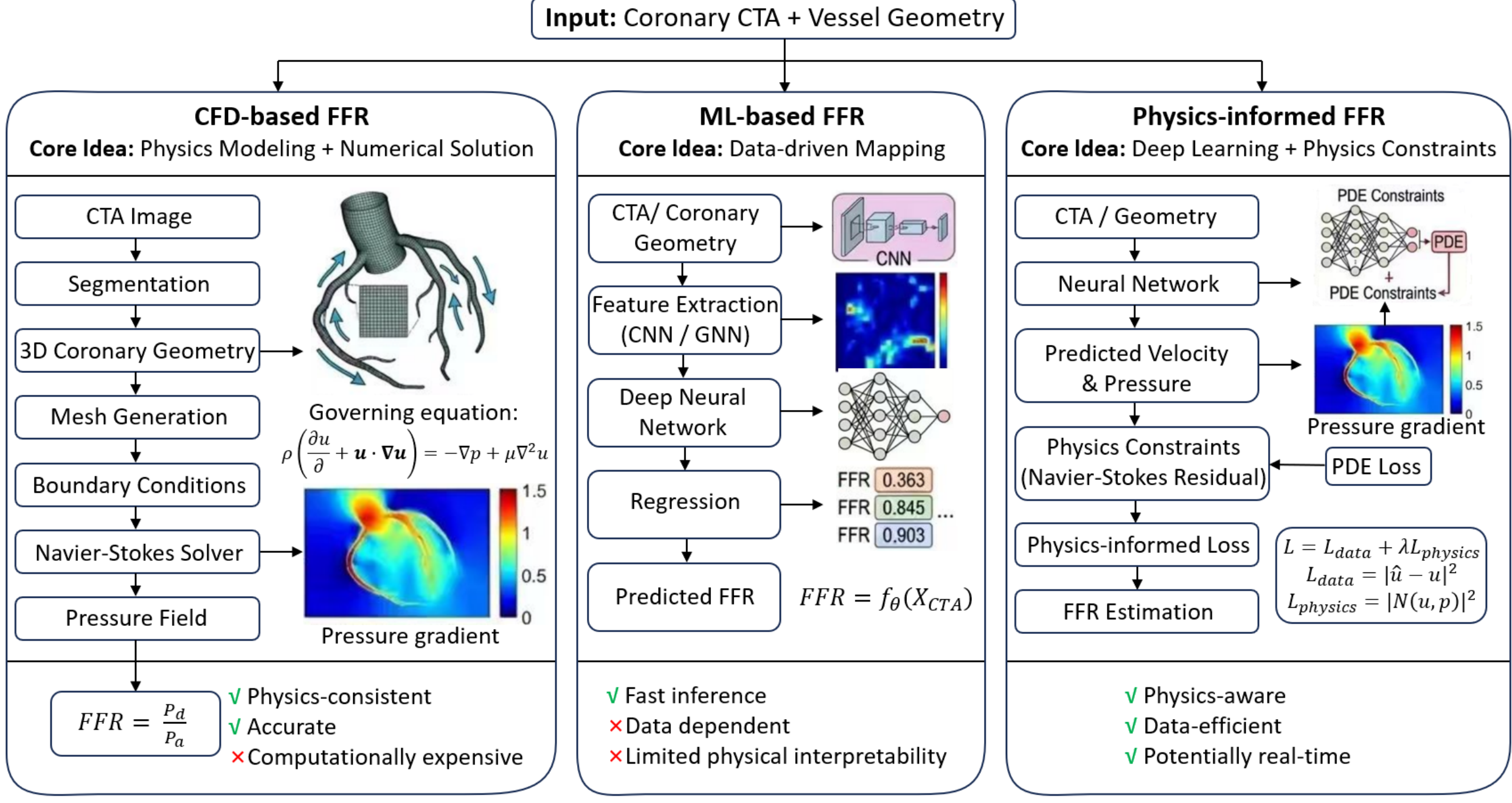

**Figure 1.** Workflow-oriented comparison of the major paradigms for imaging-derived FFR estimation, including CFD-based, machine learning–based, and physics-informed approaches. CCTA is used here as an example input. For ICA-based FFR, the downstream estimation framework is largely similar, while the initial stage typically involves ICA-based 3D coronary reconstruction rather than direct CTA-derived geometry.

### 1.4 Scope and contributions of this review

**Scope:** This review focuses on imaging-derived FFR methods for CCTA and ICA workflows, emphasizing approaches that are clinically plausible in terms of end-to-end latency and operational requirements. We cover physics-based modeling across a fidelity spectrum (3D CFD to reduced-order models) [15-18], ML/DL approaches ranging from feature-based predictors to end-to-end inference [27-30], and physics-informed learning frameworks that integrate governing-equation structure into learning [23-26, 31].

**Contributions:** We aim to provide a translation-oriented synthesis with the following contributions:

1. A workflow-grounded taxonomy spanning CCTA and angiography pipelines, mapping major methodological families to their practical trade-offs [8, 13, 14].
2. A structured overview of acceleration strategies for physics-based FFR, including reduced-order formulations and hybrid designs, and how these choices impact fidelity and feasibility [32-37].
3. A boundary-condition–centric perspective on uncertainty and robustness, highlighting how inlet/outlet assumptions and hyperemia modeling propagate into pressure drop and FFR [8, 16, 38].

4. An evaluation framework that extends beyond AUC/accuracy to include calibration, uncertainty, failure modes, quality-control gatekeeping, and computability—factors that strongly determine real-world adoption [19-22].

## 2. Physiological and Mathematical Foundations

### 2.1 Fractional Flow Reserve (FFR)

FFR is defined as the ratio of maximal hyperemic myocardial blood flow in the presence of a stenosis to the theoretical maximal hyperemic flow that would be achieved in the same artery if the stenosis were absent under the same microvascular and perfusion conditions [1, 2]. During pharmacologically induced maximal hyperemia, typically achieved with adenosine, downstream microvascular resistance is minimized and becomes relatively stable, allowing coronary flow to be approximated as proportional to pressure [39]. Mathematically, FFR can be written as:

$$FFR = \frac{Q_{stenotic}}{Q_{normal}} \quad (1)$$

where $Q_{stenotic}$ denotes the maximal hyperemic flow through a coronary artery with a stenosis, and $Q_{normal}$ denotes the hypothetical maximal hyperemic flow that the same artery would achieve if the stenosis were absent under otherwise identical conditions. Under a lumped resistance approximation, flow can be expressed as pressure drop divided by resistance [74], we can express:

$$Q_{stenotic} = \frac{P_d - P_v}{R_{stenotic}}, Q_{normal} = \frac{P_a - P_v}{R_{normal}} \quad (2)$$

where $P_d$ is the distal coronary pressure, $P_a$ is the aortic pressure, $P_v$ is the venous pressure and $R$ denotes vascular resistance. Under maximal hyperemia, the downstream microvascular resistance is minimized and assumed to remain relatively constant, so the flow ratio can be approximated by the corresponding pressure ratio. In addition, $P_v$ is typically small relative to $P_a$ and $P_d$ and is commonly neglected in the simplified derivation. As a result, FFR is approximated as the ratio of distal coronary pressure to aortic pressure under maximal hyperemia:

$$FFR = \frac{P_d}{P_a} \quad (3)$$

An FFR value of ≤0.80 is generally considered hemodynamically significant and suggests that the stenosis is likely to induce myocardial ischemia, thereby guiding revascularization decisions in clinical practice [2].

### 2.2 Governing equations and modeling abstractions (full-order to reduced-order)

FFR computation methods span a hierarchy of mathematical models, ranging from high-fidelity three-dimensional Navier-Stokes simulations to simplified reduced-order and empirical formulations. High-fidelity models provide detailed geometric representation and more complete resolution of local hemodynamics, but they require substantial computational resources and patient-specific boundary information. In contrast, reduced-order models and analytical approximations sacrifice some spatial detail in exchange for computational efficiency and improved clinical feasibility. Understanding this modeling spectrum is essential for interpreting the physical assumptions, computational trade-offs, and clinical applicability of contemporary FFR methodologies. A comparison of these models from full 3D to 0D is shown in Table 1 below.

Table 1: Model Hierarchy - Fidelity vs Computational Cost

| Model Dimension | Governing Equations | Variables | Solve Time | Fidelity | Used By |
|---|---|---|---|---|---|
| 3D | Navier-Stokes (Eqs. 4-5) | $\boldsymbol{u}, p$ | 2-4 hours | Highest | HeartFlow [9] |
| 2D | Angiography-based | $\Delta p, Q$, $U$(mean velocity) | 2 min | Medium High | FFR2D [44] |
| 1D | Area averaged (Eqs. 9-11) | $A(x,t)$ $Q(x,t)$ , $p$ | 5-10 min | High | QFR, vFFR [32, 45] |
| 0D | Algebraic resistance (Eq. 12-13) | $p, Q$ parameter $(R, C, L)$ | < 1 min | Medium | FFRangio [43] |

### 2.2.1 3D Navier-Stokes Equations

At the highest level of fidelity, coronary blood flow is commonly described by the incompressible Navier-Stokes equations. Assuming blood is modeled as a Newtonian fluid with constant density $\rho$ and dynamic viscosity $\mu$, the velocity field $\mathbf{u}$ and pressure field $p$ satisfy:

$$\nabla \cdot \boldsymbol{u} = 0 \text{ (continuity)} \tag{4}$$

$$\rho\left(\frac{\partial \boldsymbol{u}}{\partial t} + \boldsymbol{u} \cdot \nabla \boldsymbol{u}\right) = -\nabla p + \mu \nabla^2 \boldsymbol{u} \text{ (momentum)} \tag{5}$$

where $\rho \approx 1060 \text{ kg/m}^3$ is blood density, $t$ is time, $\nabla$ is the gradient operator, and $\nabla^2$ is the Laplacian operator. These coupled nonlinear partial differential equations (PDEs) are typically solved over a patient-specific three-dimensional coronary domain $\Omega$ with appropriate boundary conditions. Equation (4) enforces mass conservation, whereas Eq. (5) expresses momentum conservation through a balance among transient inertia, convective inertia, pressure gradients, and viscous diffusion.

In coronary flow analysis, the Reynolds number $Re = \rho U D/\mu$ characterizes the ratio of inertial to viscous forces, where $U$ is a characteristic velocity and $D$ is the characteristic vessel diameter. The Womersley number $\alpha = r_v\sqrt{\omega\rho/\mu}$, where $r_v$ is vessel radius and $\omega$ is the angular cardiac frequency, characterizes the importance of pulsatile inertia relative to viscous effects.

**Boundary Conditions (BC)**: Solving the Navier-Stokes equations requires specification of boundary conditions on the computational domain $\Omega$, whose boundary $\partial\Omega$ consists of an inlet surface $\Gamma_{in}$, one or more outlet surfaces $\Gamma_{out}$, and the vessel wall $\Gamma_{wall}$.

At the inlet, either a velocity boundary condition or a pressure boundary condition is prescribed:

$$\boldsymbol{u}|_{\Gamma_{in}} = \boldsymbol{u}_{\text{in}} \text{ or } p|_{\Gamma_{in}} = p_{\text{in}} \tag{6}$$

where $\mathbf{u}_{in}$ is the prescribed inlet velocity profile and $p_{in}$ is the prescribed inlet pressure at the inlet boundary $\Gamma_{in}$. For steady-state FFR computation, a parabolic velocity profile or a specified flow rate $Q_{in}$ is commonly imposed. Estimating patient-specific $Q_{in}$ remains challenging and often relies on allometric scaling relationships involving vessel size and myocardial mass [40].

The vessel wall is commonly assumed to be rigid and satisfy a no-slip condition:

$$\boldsymbol{u}|_{\Gamma_{wall}} = 0 \tag{7}$$

which neglects wall compliance and fluid–structure interaction for computational simplicity.

**Outlet BC:** Outlet boundary conditions are often the most influential and uncertain component of coronary FFR modeling because they represent the downstream microvascular circulation. Common choices range from simple prescribed-pressure conditions to lumped-parameter outlet models, such as resistance boundary conditions, Windkessel models, and coronary lumped parameter networks.

**Resistance BC**: The simplest physiologically motivated outlet model uses a resistance-type outlet boundary condition [38]:

$$\Delta p = RQ \tag{8}$$

where $Q = \int_{\Gamma_{out}} \mathbf{u} \cdot \mathbf{n} \, dA$ is the flow rate through the outlet surface, $\mathbf{n}$ is the outward unit normal vector, and $A$ is the outlet cross-sectional area. In practice, determining patient-specific resistance values is nontrivial; common approximations include Murray's law ($R \propto D^{-3}$) and allometric scaling with myocardial volume ($R \propto V_{myocardium}^{-3/4}$), where $V_{myocardium}$ donates myocardial volume.

**Windkessel Models (3-Element):** A more sophisticated outlet representation couples a proximal (or characteristic) resistance $R_c$, a distal/peripheral resistance $R_p$, and an arterial compliance $C$ [16]. This type of model captures part of the pulsatile interaction between flow and pressure, but it introduces additional parameters that are often patient-specific and difficult to measure directly.

### 2.2.2 Reduced-Order Models

To improve computational efficiency, many FFR methods employ reduced-order formulations that simplify the full three-dimensional flow field while retaining selected hemodynamic mechanisms relevant to pressure loss and coronary perfusion [18].

**1D Area-Averaged Flow Model**: The one-dimensional formulation reduces the three-dimensional vessel to an axial description obtained by averaging flow quantities over each cross-section. Let $x$denote the axial coordinate along the vessel centerline. The cross-sectional area $A(x,t)$, volumetric flow rate $Q(x,t)$, and pressure $p(x,t)$ then satisfy the following reduced conservation equations [71, 72]:

$$\frac{\partial A}{\partial t} + \frac{\partial Q}{\partial x} = 0 \text{ (continuity)} \tag{9}$$

$$\frac{\partial Q}{\partial t} + \frac{\partial}{\partial x}\left(\alpha \frac{Q^2}{A}\right) + \frac{A}{\rho}\frac{\partial p}{\partial \mathrm{x}} = -K_R \frac{Q}{A} \text{ (momentum)} \tag{10}$$

where $\alpha$ is the momentum-flux correction coefficient associated with the assumed velocity profile, and $K_R$ is a friction-related coefficient representing viscous dissipation in the reduced-order model. Compared with full 3D simulation, the 1D formulation substantially lowers computational cost by averaging flow variables over each cross-section while still preserving the dominant axial transport behavior.

A constitutive pressure–area relation is typically introduced to account for vessel wall mechanics:

$$p - p_{\mathrm{ext}} = \beta\left(\sqrt{A} - \sqrt{A_0}\right) \tag{11}$$

where $\beta$ is a wall stiffness parameter determined by vessel material and geometric properties, $p_{ext}$ is the external pressure, $A$ is the instantaneous cross-sectional area, and $A_0$ is the reference (unstressed) area. This class of 1D models can be solved much more rapidly than full 3D Navier-Stokes simulations and provides a useful reduced-order framework for fast coronary flow analysis. It also offers conceptual support for several angiography-derived FFR methods that rely on reduced-order assumptions or simplified pressure-drop formulations [41, 42].

**0D Lumped Parameter Model:** Zero-dimensional models describe the vascular system using lumped elements analogous to electrical circuits [73]. In this framework, the vasculature is represented by lumped elements, where resistance, compliance, and inertance are denoted by $R$, $C$, and $L$, respectively. Their governing relations can be written as:

$$\Delta p = RQ, Q = C\frac{d(\Delta p)}{dt}, \Delta p = L\frac{dQ}{dt} \tag{12}$$

corresponding, respectively, to resistive, compliant, and inertial effects.

For an idealized cylindrical vessel segment, these lumped parameters can be approximated as:

$$R = \frac{8\mu l_v}{\pi {r_v}^4}, C = \frac{3 l_v \pi r^3}{2Eh}, L = \frac{\rho l_v}{\pi {r_v}^4} \tag{13}$$

where $\mu$ is dynamic viscosity, $\rho$ is blood density, $r_v$ is vessel radius, $l_v$ is vessel length, $E$ is wall elastic modulus, and $h$ is wall thickness. These elements can be combined into networks such as Windkessel or coronary lumped parameter models and solved algebraically or through low-order ordinary differential equations. Although this approach is computationally efficient, it neglects detailed spatial flow structure and therefore cannot resolve local hemodynamic features explicitly.

**2D Axisymmetric CFD Formulation:** In some fluid mechanics applications, blood flow can be approximated using a two-dimensional axisymmetric model in cylindrical coordinates $(r, z)$, where $r$ and $z$ denote radial and axial coordinates, respectively. Under this simplification, incompressible flow satisfies the continuity equation:

$$\frac{\partial u_r}{\partial r} + \frac{u_r}{r} + \frac{\partial u_z}{\partial z} = 0 \tag{14}$$

and the axial momentum equation can be written as:

$$\rho\left(\frac{\partial u_z}{\partial t} + u_r\frac{\partial u_z}{\partial r} + u_z\frac{\partial u_z}{\partial z}\right) = -\frac{\partial p}{\partial z} + \mu\left[\frac{\partial^2 u_z}{\partial r^2} + \frac{\left(\frac{1}{r}\right)\partial u_z}{\partial r} + \frac{\partial^2 u_z}{\partial z^2}\right] \tag{15}$$

where $u_r$ and $u_z$ are the radial and axial velocity components. Although such axisymmetric PDE formulations are useful for idealized vessel analysis, they are less common in practical coronary FFR computation because coronary arteries are tortuous, branching, and reconstructed from limited projection views.

**Projection-Based Analytical Pressure-Drop Models:** Rather than solving full 2D axisymmetric flow equations, many angiography-based FFR methods estimate pressure loss from image-derived vessel geometry using analytical or semi-empirical pressure-drop relations [44]. A commonly used generic form is:

$$\Delta p = k_1 Q + k_2 Q^2 \tag{16}$$

where $k_1$ represents viscous friction losses and $k_2$represents nonlinear losses associated with flow separation and geometric contraction/expansion.

A more detailed formulation can explicitly incorporate stenosis geometry and temporal acceleration effects:

$$\Delta p = \left(\frac{k_v \mu}{D}\right)U + \left(\frac{k_t}{2}\right)\left[\left(\frac{A_v}{A_1}\right) - 1\right]^2 \rho|U|U + k_u \rho L_s\left(\frac{\mathrm{d}U}{\mathrm{d}t}\right) \tag{17}$$

where $k_v$ , $k_t$ , and $k_u$ are empirical coefficients associated with viscous, separation, and unsteady/inertial losses, respectively; $U$ is mean velocity; $A_v$ and $A_1$ are the reference and minimum stenotic cross-sectional areas; and $L_s$ is stenosis length. The term $|U|U$ preserves the flow direction while maintaining a quadratic dependence on flow magnitude, which is commonly used to model separation-related losses.

Single-view methods such as $\mathrm{FFR}_{\mathrm{2D}}$ may apply such relationships directly to one angiographic projection, whereas dual-view methods such as QFR, vFFR, and FFRangio rely on two or more projections to reconstruct vessel geometry and then estimate pressure loss using reduced-order or empirical formulations. Unlike traditional CFD, these approaches do not resolve the full flow field; instead, they infer pressure drop from image-derived geometric descriptors and simplified hemodynamic assumptions.

## 3. Methods for FFR Estimation

Imaging-derived FFR methods can be organized into three methodological families that differ in how they trade off fidelity, speed, interpretability, and deployment robustness. Section 3.1 reviews physics-based approaches spanning full-order CFD to fast reduced-order formulations and highlights why boundary conditions and acquisition feasibility dominate real-world performance. Section 3.2 summarizes machine learning (ML) and deep learning (DL) approaches, from physics-implicit surrogates to end-to-end models, and discusses reliability under dataset shift. Section 3.3 introduces physics-informed learning (PINNs and neural operators) as a bridging paradigm that aims to retain physical consistency while enabling rapid inference, though scalability and BC identifiability remain open challenges.

## 3.1 CFD and Physics-Based Approaches

### 3.1.1 Full-order CFD

Full-order CFD computes coronary pressure and flow by solving the Navier-Stokes equations on a patient-specific coronary geometry reconstructed from imaging. In CT-based FFR, early multicenter evidence demonstrated the feasibility and diagnostic accuracy of FFR derived from coronary CT angiography (FFR_CT) compared with invasive pressure-wire FFR, followed by the NXT trial which further validated performance in a broader cohort. Beyond diagnostic accuracy, outcome and resource-impact studies suggested that FFR_CT–guided strategies can improve care pathways and reduce unnecessary invasive testing in selected populations [4, 6].

In practice, CT-derived CFD pipelines often adopt simplified but clinically pragmatic assumptions (e.g., steady/quasi-steady hyperemic simulation, Newtonian blood, rigid walls) and rely on prescribed inlet/outlet BCs (e.g., outlet resistances/Windkessel elements) to represent hyperemic microcirculation. As a result, geometry uncertainty and BC uncertainty can directly propagate into the predicted pressure drop and FFR [8, 15, 17].

Routine clinical scalability of full-order CFD is constrained by three recurring practical barriers:

1. **Geometry and reconstruction errors:** Coronary CFD is sensitive to lumen geometry; segmentation uncertainty and CT artifacts can propagate to pressure-drop estimates. In real-world FFR_CT workflows, inadequate CT image quality can lead to analysis rejection, and motion-related degradation is a common contributor [7, 8]. Calcium/stent blooming remains a well-recognized limitation for lumen delineation and stenosis characterization on cardiac CT [9, 10].
2. **Boundary-condition uncertainty and physiologic assumptions:** Patient-specific inflow, outlet resistance allocation across branches, and the degree of hyperemia are rarely directly observed in routine imaging workflows; plausible BC choices can yield different pressure–flow solutions, especially in complex branching trees. This sensitivity is a primary driver of inter-study variability and motivates the use of patient-specific surrogates and uncertainty-aware reporting (expanded in Section 3.1.3).
3. **Computational burden:** Full-order CFD typically requires time-consuming meshing, solver iterations, and quality control, limiting turnaround time and broad deployment in routine care [3].

To reduce runtime while retaining explicit flow physics, two complementary directions are common. First, algorithmic acceleration keeps 3D CFD but dramatically reduces compute time; in angiography-derived workflows, a fast pseudo-transient strategy reported vFFR computation in 189 s (~3.2 min) compared with >24 h for full-transient 3D CFD, representing a >450× reduction in computation time while maintaining clinically relevant agreement in its evaluation setting [37].

Second, physics-preserving model reduction reduces dimensionality to improve practicality: 1D mass–momentum equations are obtained by cross-sectional averaging along the vessel centerline and are often coupled to 0D elements for downstream microvascular beds [9]. Extensions include probabilistic 1D formulations to quantify uncertainty and tune model parameters (e.g., resistance/stenosis terms) using uncertainty quantification and Bayesian optimization [33].

These constraints motivate the broader class of fast-physics and reduced-order methods designed to fit workstation or cath-lab time budgets (Section 3.1.2).

### 3.1.2 Fast physics and reduced-order formulations for imaging-derived FFR

Reduced-order physics aims to preserve essential hemodynamics while avoiding full 3D transient CFD. In angiography-based workflows, a representative pattern is 3D-QCA reconstruction plus practical flow/velocity estimation, enabling minutes-scale computation and supporting "wire-free" physiology in the cath lab. Early angiography-derived CFD variants demonstrated feasibility from angiography alone [35], and subsequent pipelines combined 3D-QCA with contrast-flow surrogates (e.g., Thrombolysis In Myocardial Infarction [TIMI] frame count) to enable more practical computation [36]. Further acceleration using paired steady-state or pseudo-transient strategies reduces runtime while retaining explicit flow physics [37]. More recent hybrid strategies anchor CFD using invasive pressure measurements while using angiography-derived geometry, enabling estimation of intracoronary flow/resistance surrogates and evaluation against functional imaging such as SPECT myocardial perfusion imaging [46].

Another widely used reduced-order direction is 1D/0D physics, where cross-sectionally averaged equations (often coupled to lumped microvascular models) enable rapid whole-tree analysis while retaining governing-equation structure [9]. Such formulations are also conceptually aligned with several angiography-derived physiology techniques (e.g., QFR/vFFR families) that rely on simplified pressure–flow modeling to estimate lesion-level physiology under practical acquisition constraints.

More recently, 2D fluid-mechanics approaches have been proposed to reduce reliance on full 3D reconstruction by operating more directly on 2D angiograms [44]. While promising for speed and workflow integration, such approaches still require careful external validation and clear characterization of when 2D assumptions hold.

**Practical feasibility and failure modes:** Across fast-physics methods—especially those relying on 3D-QCA—real-world performance is often limited by the quality of angiographic acquisition and the feasibility of reconstruction. Common failure modes include:

1. Foreshortening, vessel overlap, panning, and suboptimal contrast filling, which can prevent reliable reconstruction or degrade downstream estimation.
2. Insufficient projection separation / inadequate views, which limit 3D geometry recovery and flow estimation in multi-view pipelines.

3. Violation of physiologic assumptions (e.g., imperfect hyperemia equivalence or microvascular dysfunction), which can degrade diagnostic accuracy and motivate explicit quality control and appropriate case selection [13, 14].

### 3.1.3 Boundary conditions

Across both full-order CFD and reduced-order physics, uncertainty in image-based FFR computation is often dominated by the specification of BCs. The most influential sources include:

1. Inlet conditions, where inflow or inlet pressure is typically inferred from imaging-derived surrogates or population-based priors.
2. Outlet conditions, particularly the allocation of distal resistance—and thus flow split—across a branching coronary tree.
3. Hyperemia modeling, since the intended state of maximal vasodilation varies across patients and may be substantially attenuated in microvascular dysfunction.

Together, these assumptions shape the pressure–flow relationship and can lead to meaningful shifts in predicted pressure drops and FFR values. Clinical guidance therefore emphasizes that FFR_CT interpretation should be integrated with anatomic CTA findings and broader clinical context, recognizing that modeling assumptions and patient-specific physiological conditions can influence the computed result [3].

For angiography-derived approaches, feasibility constraints—projection geometry, foreshortening, overlap, and contrast filling—interact with physiological assumptions and reconstruction/flow-estimation steps, amplifying error and non-computability in real-world settings. This reinforces the need for explicit quality control and, where available, incorporation of patient-specific physiological surrogates to reduce reliance on fixed priors and improve robustness across heterogeneous populations and acquisition conditions [13, 14]. Representative milestones in CFD-based imaging-derived FFR are summarized in Table 2.

Table 2. CFD-based FFR timeline

| Year | Method | Image modality | Key optimization | Dataset | Results |
|---|---|---|---|---|---|
| 2011 | CT-derived 3D CFD FFR (FFR_CT) — DISCOVER-FLOW [4] | CCTA | First end-to-end CCTA→3D CFD→FFR clinical feasibility | 103 patients 159 vessels from 3 centers | AUC 0.90 ACC 0.843 SEN 0.879 SPE 0.822 Correlation r = 0.72 mean difference -0.022±0.116 |
| 2013 | Angio-derived CFD vFFR from | ICA | Established angiography→3D | 19 patients 35 vessels | ACC 0.97 SEN 0.86 SPE 1.00 |

| | angiography alone [35] | | recon→CFD→vFFR paradigm | | PPV 1.00<br>NPV 0.97<br>Correlation r=0.84 |
|---|---|---|---|---|---|
| 2014 | Angio-derived CFD with 3D-QCA + TIMI frame count [36] | ICA | Practical flow/velocity estimation to enable faster cath-lab computation | 68 patients<br>77 vessels | AUC 0.93<br>ACC 0.88<br>SEN 0.78<br>SPE 0.93<br>PPV 0.82<br>NPV 0.91<br>Computation <10 min |
| 2017 | Minutes-scale vFFR via paired steady-state / “pseudo-transient” CFD [37] | ICA | Replaced full transient CFD with fast steady-state strategy | 73 vessels | SEN 1.00<br>SPE 1.00<br>Computation time 189s (~3.2 min) vs >24h for full transient<br><1% error vs transient CFD |
| 2018 | Physics-preserving 1D NS (+0D coupling) benchmarked vs 3D [9] | CCTA | Orders-of-magnitude speedup while retaining governing-equation structure | 20 patients<br>29 geometries | AUC 0.97<br>ACC 0.98<br>SEN 0.99<br>SPE 0.90<br>1D model ~1000× faster |
| 2019 | UQ + Bayesian optimization for 1D coronary FFR. [33] | CCTA | Explicit inference/calibration of uncertain BC/parameters | 11 patients<br>22 vessels | Reduced parameter uncertainty by 0.6-0.8<br>Computation time ~10 min per patient Validated against 3D CFD |
| 2024 | 2D FFR (2D angiography fluid-mechanics) [44] | ICA | Reduced reliance on full 3D recon by using 2D angiograms | 88 patients | ACC 0.909<br>SEN 0.857<br>SPE 0.933<br>Correlation r=0.68<br>Computation 0.1s |

### 3.2 Machine Learning Approaches

Machine-learning approaches complement physics-based FFR estimation by replacing part (or all) of the CFD pipeline with data-driven inference. The overarching promise is scalability—near-instant computation and easier deployment—while the main clinical requirement is reliability under heterogeneous scanners, protocols, image quality, and lesion morphologies.

#### 3.2.1 ML surrogates of CFD (physics-implicit)

Physics-implicit surrogates learn a direct mapping from anatomy-derived representations (e.g., centerline geometry, lumen-area profiles, stenosis descriptors, vessel-tree/graph features, or MPR-derived profiles) to pressure drop or FFR using CFD outputs and/or invasive FFR supervision. Early "classical ML" CT-FFR studies established workstation-feasible inference and multi-center validation before modern end-to-end DL became mainstream [28, 29]. In parallel, traditional supervised ML using quantitative angiographic descriptors provided early baselines for predicting FFR-defined ischemia in intermediate lesions [30].

Clinically oriented ML-based CT-FFR has been reported to outperform anatomy-only CCTA for identifying hemodynamically significant disease [27]. The field has increasingly emphasized on-site/workstation pipelines: a high-speed on-site DL-FFR-CT workflow reported AUC 0.975, ACC 95.9%, SEN 93.5%, SPE 97.7% (FFR-CT $\leq$ 0.80) with strong agreement with invasive FFR ($r \approx 0.81$) [47]. A 2025 SMART trial summary similarly reported strong vessel-level discrimination (AUC $\approx$ 0.95) with high operating performance and near–real-time computation [48]. At the evidence-synthesis level, meta-analyses suggest ML-FFRCT and CFD-FFRCT have broadly comparable discrimination/specificity, but ML approaches may exhibit lower sensitivity in some aggregates, implying a practical trade-off between speed and potential missed ischemic lesions without careful calibration/external validation [49, 50].

#### 3.2.2 End-to-end deep learning

DL models learn directly from CCTA volumes or artery-centered MPR stacks (3D CNN or CNN–attention hybrids) and may incorporate anatomy-aware vessel/tree representations to stabilize regression. In a multi-center evaluation, Hampe et al. reported AUCs of 0.78 (merged output) and 0.83 (regression head) for predicting invasive FFR from CCTA, highlighting both feasibility and sensitivity to protocol heterogeneity [51]. Beyond Transformers, lightweight attention blocks such as SE/CBAM are widely used to highlight lesion-relevant features and suppress artifact-dominated regions [52, 53]. A 2025 onsite CCTA-FFR study using invasive iFR as reference reported AUC 0.79 with ACC 81.8% (SEN 89.3%, SPE 68.8%), illustrating that onsite practicality is improving while performance remains cohort- and endpoint-dependent [54]. In parallel, feature-based ML using quantitative plaque/CTA descriptors remains relevant as an interpretable alternative, linking imaging biomarkers to FFR-defined ischemia and impaired myocardial flow [55].

Angiography DL models range from single-frame predictors to video-based models that exploit contrast-flow dynamics, sometimes aiming for AI-only inference without explicit 3D reconstruction/CFD. Related work has shown that coronary flow velocity fields can be inferred from fluid-component concentration with high-performance neural networks, highlighting the value of contrast dynamics for field-level learning [56]. Temporal learning is commonly implemented via 3D CNNs or sequence models such as ConvLSTM/TCN [57, 58]. A single-frame transfer-learning baseline reported AUC/ACC ~0.81 for FFR≤0.80 classification in intermediate LAD lesions [59]. Video-aware models can achieve substantially higher discrimination in their test settings: Mineo et al. reported AUC 0.95±0.03 (single-view, 3D) and 0.95±0.04 (multi-view, 3D) with ACC ~89% (SEN ~80%, SPE ~96–99%) [60]. A multi-center evaluation of an AI-only angiography-derived FFR system reported AUC 0.93 with ACC 93.7%, while highlighting vulnerability to acquisition/domain shifts (frame rate, injection, vendor post-processing, projections/overlap) [61]. Finally, physics-guided/tree-aware DL that predicts pressure/FFR fields from vessel-tree representations reflects a trend from single-index regression toward whole-tree functional mapping, potentially enabling richer physiological outputs than a single lesion-level number [62].

More broadly, machine learning on invasive coronary angiography has also been extensively studied for arterial extraction, major-vessel semantic labeling, stenosis evaluation, and ICA-based FFR estimation, underscoring that robust upstream angiographic representation is an important foundation for downstream functional prediction [63].

#### 3.2.3 Uncertainty, calibration, and domain generalization

Because ML-FFR informs high-stakes decisions (PCI vs. deferral), point estimates should ideally be accompanied by calibrated confidence intervals and/or uncertainty estimates. Neural networks can be substantially mis-calibrated; temperature scaling is a standard post-hoc method that can reduce miscalibration [19]. However, calibration and uncertainty quality often deteriorate under dataset shift; large-scale studies show that i.i.d. calibration does not guarantee calibration under corruptions or domain changes, and ensembles are often among the most robust approximations [21]. Uncertainty is commonly framed as epistemic vs aleatoric (e.g., Bayesian approximations or ensembles), supporting selective prediction and risk-aware decision support [20].

A practical pattern is uncertainty-guided triage: uncertain/borderline cases (poor image quality, heavy artifacts, atypical anatomy, protocol mismatch) are flagged for invasive FFR/iFR or expert review rather than forcing an automated decision. This concept has been demonstrated explicitly; for example, in the CT DL study above, "correcting" the top 20% most-uncertain cases to reference values improved performance to AUC 0.91, ACC 0.89, SEN 0.92, SPE 0.78 (vs all-artery AUC 0.78 / ACC 0.79) [51]. More broadly, domain generalization provides a toolbox to improve out-of-distribution robustness across centers and protocols, motivating shift-aware validation (external sites, temporal splits), routine reporting of calibration (ECE/reliability curves), and conservative referral policies when confidence is low [22]. Representative milestones in ML-based imaging-derived FFR are summarized in Table 3.

Table 3. ML-based FFR timeline

| Year | Method | Image modality | Key optimization | Dataset | Results |
|---|---|---|---|---|---|
| 2018 | Workstation-feasible ML-based CT-FFR (feature-based / classical ML) [28, 29] | CCTA | Replaced iterative CFD with learned surrogates; enabled fast | 254 patients | AUC 0.89<br>SEN 0.79<br>SPE 0.94<br>Computation 2.4s |
| | | | inference from CCTA-derived quantitative features (pre end-to-end DL era) | | vs 10 min for CFD Equivalent accuracy to CFD-FFR |
| 2019 | Angiography feature-based ML for FFR classification [30] | ICA | Introduced quantitative angiographic descriptors → supervised ML baseline for predicting FFR≤0.80 (pre video-DL) | 150 patients | AUC 0.84±0.03 ACC 0.78±0.04 With 12 top features: AUC 0.87±0.01 ACC 0.81±0.01 SEN 0.84 SPE 0.80 |
| 2022 | Multi-center CT DL for invasive-FFR prediction [51] | CCTA | Scaled to heterogeneous protocols; 3D CNN / artery-centered inputs + regression heads; highlighted generalization limits under multi-center shift | 569 patients from 3 hospitals 514 arteries with FFR | AUC 0.78 ACC 0.79 SEN 0.84 SPE 0.61 Patient-level AUC 0.75 Accuracy 0.80 |
| 2023 | ML-based CT-FFR clinical evidence / positioning [27] | CCTA | Strengthened clinical evidence that ML-CTFFR improves over anatomy-only CCTA for ischemia detection | Meta-analysis and clinical studies | AUC 0.86 SPE 0.80 SEN 0.86 Positive LR 5.8 Higher diagnostic performance than CCTA alone |
| 2023 | High-speed on-site DL-FFR-CT [47] | CCTA | On-site, near–real-time inference; reduced turnaround time and dependence on off-site pipelines while maintaining high | Validation cohort (prospective) | AUC 0.975 ACC 0.959 SEN 0.935 SPE 0.977 Correlation r≈ 0.81 Computation time ~7 min 54s per patient |

| | | | diagnostic performance | | |
|---|---|---|---|---|---|
| 2023 | AI-only angiography-derived FFR (multi-center) [61] | ICA | End-to-end angiography → FFR without explicit CFD/3D recon; enabled fast, wire-free functional assessment (but domain-shift sensitive) | 297 patients<br>304 vessels from 3 centers | AUC 0.93<br>ACC 0.94<br>SEN 0.91<br>SPE 0.95<br>Calculation time 37.5±7.4s per video<br>Automatic lesion localization 0.89 success |
| 2024 | Spatiotemporal/video DL for angiography FFR/iFR [60] | ICA | Modeled contrast-flow dynamics using 3D CNN + self-attention; major jump from single-frame to sequence-aware inference | 389 patients<br>778 angiography exams | AUC 0.95±0.03 (single-view)<br>0.95±0.04 (multi-view)<br>ACC 0.89<br>SEN 0.80<br>SPE 0.96-0.99<br>No key-frame selection needed |
| 2024 | Single-frame angiography DL baseline (intermediate lesions) [58] | ICA | Established practical image-only baseline (screening-level), clarifying the incremental value of temporal modeling | 41 patients<br>3625 images ( LAD lesions) | AUC 0.81<br>ACC 0.81<br>SEN 0.86<br>SPE 0.75<br>F1 0.84 (DenseNet169)<br>Transfer learning from ImageNet |
| 2025 | DL CTA-FFR trial/clinical workflow consolidation (SMART summary) [48] | CCTA | Reinforced deployability trend: workstation-integrated, seconds-scale computation with strong vessel-level discrimination | 339 patients<br>414 vessels | Per-vessel:<br>SEN 0.947<br>SPE 0.886<br>ACC 0.908<br>AUC 0.95<br>Per-patient:<br>SEN 0.938<br>SPE 0.88<br>ACC 0.903<br>Computation 22.5±1.9s |
| 2025 | Physics-guided / vessel-tree–aware DL | CCTA | Shift from single-number | 273 patients | AUC 0.90<br>SEN 0.984 |

| | (whole-tree functional mapping) [62] | | regression to pressure/FFR field prediction on coronary trees; improved long-range dependency modeling via attention | | SPE 0.801<br>ACC 88.2%<br>Near real-time inference (<1s) |
|---|---|---|---|---|---|

### 3.3 Emerging Direction: Physics-Informed Learning

Physics-informed learning has emerged as a practical bridge between CFD and ML for imaging-derived FFR by embedding the structure of governing equations (e.g., conservation laws and boundary consistency) into learning and inference. As discussed above, full-order CFD is physiologically grounded but often limited by meshing/solver burden and sensitivity to geometry and boundary-condition assumptions, whereas purely data-driven ML can be fast yet vulnerable to domain shift and non-physiological predictions when confronted with out-of-distribution anatomy or acquisition protocols. Physics-informed approaches aim to mitigate these complementary limitations by coupling learned representations with explicit physical constraints. The general PINN formulation for PDE-constrained learning is well established [23], with fluid-mechanics–focused guidance on training practices and failure modes [24]. Recent FFR-oriented studies increasingly report quantitative agreement with measured or reference FFR while aiming for fast inference and improved robustness across anatomy and acquisition conditions.

#### 3.3.1 Classic PINNs and challenges

Classic PINNs (i.e., pointwise networks trained by minimizing PDE residuals plus boundary/initial-condition losses) have been directly applied to coronary hemodynamics as a surrogate to recover pressure/flow fields and derive FFR-related endpoints, typically by learning a continuous 3D pressure–velocity field in the coronary tree under Navier-Stokes constraints without explicit meshing/iterative CFD solving. Representative evidence includes a 3D PINN-based coronary-tree study reporting mean error versus invasive FFR of approximately 1.2%, 2.3%, and 2.8% on three evaluated arteries/cases [64]. These results support the feasibility of patient-specific modeling when physical constraints are correctly enforced, and boundary conditions are well-posed.

Despite this promise, coronary FFR remains challenging for classic PINNs. Coronary geometries are complex (tortuosity, bifurcations, diffuse disease), and solutions can be highly sensitive to stenosis morphology and boundary conditions. Training can be stiff and unstable when enforcing PDE residuals and boundary constraints over irregular domains, with convergence and reproducibility varying across cases [24]. Methodological work in vascular hemodynamics suggests that optimization stability can be improved by training strategies such as progressively increasing boundary-condition complexity [31]. Meanwhile, boundary-condition identifiability remains a core bottleneck—multiple outlet-resistance/hyperemia configurations may yield physically plausible pressure fields, motivating stronger priors, auxiliary measurements, and uncertainty-aware validation before clinical deployment.

### 3.3.2 Beyond classic PINNs: Neural operators and hybrid frameworks

To improve scalability across heterogeneous anatomies and boundary conditions, neural operator learning (e.g., FNO) learns mappings between function spaces and can generalize across PDE families [25]. PINO further injects PDE residual constraints into operator learning to improve data efficiency and physical fidelity while preserving rapid inference [26]. In FFR-specific applications, recent work illustrates multiple "physics-informed hybrid" routes with measurable performance:

**Physics-informed graph learning:** A conditional physics-informed graph neural network was evaluated on 183 real-world coronaries (143 X-ray, 40 CT angiography), reporting a strong correlation between predicted and measured FFR (r = 0.89 on X-ray and r = 0.88 on CT) [65].

**Physics-informed neural operator for angiography-derived FFR curves (BVPINO):** In coronary angiographies from 215 vessels / 184 subjects, BVPINO was reported to achieve an "optimal balance" of effectiveness and efficiency, with high agreement/correlation between distal FFR predictions and invasive FFR. BVPINO achieves near real-time inference (< 1s) while maintaining competitive diagnostic performance. For ischemia detection, it reports AUC = 0.960 and reaches ACC = 0.916 / F1 = 0.932 at the clinical threshold (FFR $\leq$ 0.80); for disease-pattern identification, it reports AUC = 0.978 (diffuse) and 0.969 (focal) with ACC = 0.861 / F1 = 0.846 (PPG-based). Computation-based CFD and PINN baselines achieve slightly higher ACC/F1 but at much higher runtime (200 s for CFD; 480 s for PINN): ischemia (FFR) ACC/F1 = 0.921/0.939 (CFD) and 0.926/0.941 (PINN) vs 0.916/0.932 (BVPINO), and disease pattern (PPG) ACC/F1 = 0.867/0.850 (CFD) and 0.873/0.852 (PINN) vs 0.861/0.846 (BVPINO), highlighting a strong accuracy–efficiency trade-off for cath-lab deployment [66].

**Physics-informed self-supervised digital-twin learning:** PINS-CAD pretrains GNNs on 200,000 synthetic coronary digital twins with 1D Navier-Stokes/pressure-drop constraints and fine-tunes on 635 FAME2 patients, achieving AUC = 0.73 for predicting future cardiovascular events; it also generates spatially resolved pressure and FFR curves along centerlines [67].

Overall, these results suggest a consistent trend: physics-informed constraints can reach high correlation / competitive accuracy in selected datasets while enabling fast inference and richer outputs (e.g., vessel-wise FFR curves).

### 3.3.3 Outlook: likely roles in near-term clinical translation

Near-term translation is most plausible via hybrid workflows: (i) using physics-informed losses/structures to regularize purely data-driven predictors; (ii) deploying physics-informed models as fast, constraint-aware surrogates for pressure/FFR mapping; and (iii) enabling risk-aware reporting (sensitivity or confidence flags) when boundary conditions are uncertain. The field is moving quickly, but broader external validation and clearer reporting of standardized endpoints (AUC/MAE/r, plus failure-case analyses) remain essential.

## 4. Evaluation and Clinical Translation

This section shifts the focus from "how high is the average AUC/accuracy" (covered in Section 3) to questions that determine real-world adoption: (i) can the method be computed reliably in routine workflows, (ii) what are the dominant failure modes and how should QC gatekeeping be implemented, (iii) how do study design choices introduce bias, and (iv) how do complex lesion morphologies (e.g., bifurcation and diffuse disease) affect feasibility and clinical usefulness.

### 4.1 Workflow integration, quality control, and failure modes

In practice, the main bottleneck is often not model discrimination but workflow readiness: computational latency, user interaction burden, input quality requirements, and robust handling of failure. Here, quality control (QC) refers to explicit criteria used to decide whether an exam is analyzable/reportable (e.g., adequate image quality and acquisition requirements), and to trigger fallback strategies (repeat acquisition, alternative functional testing, or invasive pressure-wire FFR).

**CT-FFR:** Common failure modes are closely linked to CCTA quality and segmentation/reconstruction robustness, including motion artifacts, noise/poor contrast timing, heavy calcification with blooming, and other cardiac CT artifacts that degrade lumen delineation [8-10]. In a large real-world analysis of technically unsuccessful CT-derived FFR, the FFR_CT rejection rate due to inadequate image quality was 2.9% in the ADVANCE registry, but increased to 8.4% in a consecutive clinical cohort. Motion artifacts were the dominant reason for failure in both cohorts (63/80, 78% in ADVANCE; 729/892, 64% in the clinical cohort) [5]. These data emphasize that computability/coverage and structured failure reporting should be treated as primary translation endpoints rather than secondary footnotes.

Workflow speed is improving through algorithmic acceleration and learned surrogates. For example, a high-speed on-site deep learning–based FFR-CT workflow reported a mean analysis time of 7 minutes 54 seconds per patient, moving closer to clinical decision timelines [47]. However, faster inference does not eliminate the need for QC: low-quality inputs can still produce unreliable results or increase rejection rates, so deployment requires a clear gatekeeping policy and a safe fallback pathway.

**Angiography-derived physiology:** Feasibility depends strongly on acquisition geometry and 3D-QCA/3D reconstruction quality. In a real-world “all-comers” QFR vs FFR study, QFR was attempted in all cases and was successful in 73% of patients (568/778 vessels). Among 210 failed vessel-level attempts, the most frequent cause was the absence of two projections with ≥25° separation (43%), followed by excessive vessel overlap (31%), insufficient contrast filling (16%), and inadequate image quality (10%). Importantly, QFR measurement time and accuracy showed a clear learning curve: median measurement time was 271 s (IQR 206–374), decreasing from 560 s (IQR 418–719) in the first 100 calculations to 226 s (IQR 168–251) in the last 100; in parallel, median absolute bias improved from 0.05 to 0.03 [14]. These results quantify the operator- and workflow-dependence of the technique and motivate standardized acquisition /QC protocols.

More automated, AI-based angio-FFR systems can further reduce per-case latency but still require user interaction in a nontrivial fraction of cases. In one multi-center evaluation of an AI-based angiography-derived FFR approach, calculation time was 37.5 ± 7.4 s per angiographic video; the software automatically located the target lesion in 89% of cases, while 11% required manual lesion marking by the operator [61]. In a prospective multicenter study of an angiography-derived tool (AccuFFRangio), the total operating time was 4.20 ± 1.53 minutes, supporting feasibility within a catheter-lab workflow [68]. Overall, translation requires reporting not only accuracy but also end-to-end time, human-in-the-loop frequency, QC pass/fail rates, and structured failure modes with explicit fallback policies. Key non-computability rates and dominant failure modes across modalities are summarized in Table 4.

Table 4. Non-computability and dominant failure modes

| Modality | Non-computable / excluded | Dominant reasons |
|---|---|---|
| CT-FFR | 2.9% (80/2778) to 8.4% (892/10,621) rejected [5] | Motion artifacts were the leading cause (63/80; 729/892) |
| QFR | Failed in 210 vessels (out of 778 attempts) [14] | Missing ≥2 projections with ≥25° separation (90, 43%); overlap (66, 31%); poor contrast (34, 16%); poor image quality (20, 10%) |
| vFFR (FAST II) | 54/391 excluded due to unsuitable angiograms; +3 excluded for analysis reasons [41] | Exclusion causes include overlap, table movement, and foreshortening |
| AI angio-FFR | Manual lesion marking needed in 11% [61] | Automatic lesion localization succeeded in 89% |
| AccuFFRangio | On-site analysis was successful in all 304 interrogated vessels (analysis cohort) [68] | Upstream exclusions included FFR measurement failure (8 patients, 2.5%) and other reasons |

### 4.2 Study design and bias (selection bias, reference standards)

Selection bias and "computability bias." Many validation studies implicitly enrich for analyzable images by excluding cases with poor acquisition, heavy overlap, or insufficient projections. This inflates reported accuracy relative to real-world practice. A concrete illustration is the gap between controlled acquisition studies and all-comers practice: online QFR and invasive FFR were both obtained in 328/332 vessels (98.8%) in a prospective multicenter study [11], whereas real-world all-comers evidence shows substantially lower computability with angiographic acquisition–driven failures (see Section 4.1) [14]. Similarly, vFFR validation explicitly excluded a non-trivial fraction of enrolled patients for angiographic reasons (~14% in FAST II) [41], and CT-FFR feasibility also differs meaningfully between curated settings and routine consecutive cohorts due to image-quality limitations (see Section 4.1) [5]. For fair clinical translation, intention-to-diagnose reporting (including non-computable cases) is therefore essential.

**Reference standards and endpoint definition:** Most studies use pressure-wire FFR (≤0.80) as the primary reference standard, but reference measurements themselves have procedural variability (e.g., hyperemia quality, pressure drift, wire position), which can blur the "ground truth" and should be standardized and reported [3]. Moreover, because different modalities estimate physiology under different assumptions (e.g., flow/velocity surrogates, microvascular modeling), subgroup analyses (acute vs chronic presentations; lesion length/diameter; gray-zone physiology) are important to avoid spectrum bias [68]. When possible, analyses should distinguish patient-level from vessel-level endpoints and should clarify handling of multivessel disease and lesion selection rules.

**Bias controls that should be explicitly documented:** Translation-focused studies should (i) pre-specify inclusion/exclusion and QC criteria, (ii) report the fraction and causes of non-computability, (iii) ensure blinded core-lab vs on-site comparisons where relevant, and (iv) provide sensitivity analyses that treat non-computable outputs as test failures rather than removing them post hoc [7, 14, 41].

### 4.3 Metrics and endpoints: from diagnostic accuracy to clinical utility

**Representative diagnostic benchmarks:** Landmark CT-FFR trials reported strong discrimination beyond anatomy-only CCTA, with per-vessel AUC around 0.90 and accuracy in the low-to-mid 80% range in early validation (e.g., DISCOVER-FLOW) [4], while DeFACTO reported per-patient AUC of 0.81 with high sensitivity (90%) but modest specificity (54%), highlighting the practical importance of false positives [6]. Subsequent prospective multicenter validation (e.g., NXT) improved overall diagnostic performance (per-patient AUC ~0.90; accuracy ~81%) under more standardized protocols [5]. For angiography-derived methods, online QFR achieved vessel-level diagnostic accuracy of 92.7% (304/328) [11], vFFR reported diagnostic accuracy of 90% (sensitivity 81%, specificity 95%) in FAST II [41], and on-site AccuFFRangio achieved accuracy of 95.07% with AUC 0.972 in a prospective multicenter study [68]. These multi-modality numbers serve as useful anchors, but translation requires more than AUC.

Agreement, reproducibility, and "beyond-threshold" endpoints. Many clinical decisions depend on how close estimates are to the wire-based reference, especially in gray-zone physiology. In real-world QFR–FFR comparisons, the median absolute difference was 0.03, and 7% of cases differed by >0.10 (QFR vs FFR), which can meaningfully change downstream treatment decisions [14]. In ACCURATE, AccuFFRangio showed a mean difference of $0.01 \pm 0.06$ versus FFR and high discrimination (AUC 0.972) [68]. Reproducibility is also a deployment requirement: an on-site DL-FFR-CT workflow reported interobserver ICC of 0.94 and intraobserver ICC of 0.85, suggesting that the end-to-end pipeline can be consistent when executed under standardized conditions [47]. For diffuse or serial disease, pullback-shape agreement (not only distal-point classification) becomes particularly relevant, because a single distal value may not capture distributed pressure loss.

**Clinical impact endpoints:** Translation should ultimately show patient-management benefit rather than only diagnostic performance. In PLATFORM, an FFR-CT–guided strategy reduced unnecessary invasive angiography in the "planned invasive" cohort; among those who proceeded to angiography, the proportion without obstructive CAD was 12% versus 73% in usual care, consistent with improved gatekeeping [6]. For angiography-based physiology, FAVOR III China reported improved 1-year clinical outcomes with QFR-guided lesion selection (MACE 5.8% vs 8.8%; hazard ratio 0.65) [12]. These outcome-oriented endpoints provide higher-level evidence that physiology-informed strategies can change care pathways, provided feasibility and QC are addressed.

### 4.4 Lesion location and morphology: bifurcation and diffuse disease

Lesion morphology and location can strongly affect both feasibility and clinical usefulness and should be treated as explicit stratification variables in translation studies.

**Bifurcation, overlap, and complex geometry:** Angiography-derived methods are sensitive to projection requirements and overlap, which are often more challenging in bifurcation and tortuous anatomy. Real-world all-comers evidence indicates that overlap/foreshortening and inadequate projection separation are major drivers of non-computability (see Section 4.1), consistent with practical QFR requirements and head-to-head studies of performance determinants [13, 14]. For deployment, bifurcation-specific QC checks (overlap, projection separation) should be explicit, with standardized guidance for re-acquisition when QC fails.

**Diffuse disease and serial lesions:** Diffuse atherosclerosis produces distributed pressure loss without a single focal stenosis, making a binary "FFR ≤0.80 at one point" an incomplete descriptor of physiology. FAST II explicitly defined diffuse disease as wall irregularities without focal lesions with intermediate diameter stenosis (30–70%), underscoring that physiologic assessment often extends beyond discrete stenosis grading [41]. For such morphologies, pullback-style endpoints (shape and total pressure-loss distribution) and robustness to microvascular assumptions become more important than point estimates alone.

**Severe calcification and CT blooming:** For CT-FFR, heavy calcification and blooming artifacts can degrade lumen assessment and increase technical failure or bias; artifact mechanisms and mitigation strategies are well described in cardiac CT literature [9, 10]. Real-world experience indicates that challenging morphologies frequently coexist with challenging acquisition conditions and higher failure risk (see Section 4.1), reinforcing the need for QC gatekeeping and robust fallback strategies [5].

Overall, translation studies should move beyond pooled performance and explicitly report feasibility, failure modes, and decision-impact metrics across lesion morphologies (bifurcation vs non-bifurcation; focal vs diffuse; calcified vs non-calcified), because these strata often determine where imaging-derived physiology provides the greatest clinical value—and where it is most likely to fail without rigorous QC.

## 5. Conclusion

Imaging-derived FFR research is moving from optimizing AUC alone toward translation-driven goals: clinical time budgets, scalability, and robustness across centers. Between full-order CFD (mechanistically grounded but slow and workflow-heavy) and purely data-driven learning (fast but potentially non-physiological under domain shift), physics-informed learning, especially PINNs and neural operators (FNO/PINO), is emerging as a practical bridge. It aims to preserve key physical structure while enabling near real-time inference and richer outputs, such as centerline pressure/FFR curves.

Classic PINNs learn pressure–velocity fields by minimizing PDE residuals pointwise, but training can be unstable in complex coronary trees and sensitive to how boundary conditions are posed. In contrast, neural operators better align with clinical-scale deployment by learning function-to-function mappings that generalize across geometries and boundary configurations. PINO further injects PDE residual constraints into operator learning, improving physical fidelity without sacrificing speed. Importantly, operator-based frameworks naturally support curve-level physiology, not only a single distal value—an advantage for diffuse or serial disease where pullback-like shape and distributed pressure loss are clinically meaningful. Recent FFR-oriented evidence also suggests that PINO-style models can achieve near real-time inference with competitive agreement and ischemia-detection performance, offering a compelling accuracy–efficiency trade-off for cath-lab workflows.

From a clinical translation and evaluation perspective, these methodological advantages matter only if studies address workflow questions in real practice: Can the method be computed reliably? How long does it take end-to-end? How much human interaction is required? What happens when computation fails? Accordingly, beyond AUC/accuracy, translation-focused evaluations should routinely report, under an intention-to-diagnose framework, coverage/non-computability rates and dominant causes (CT: motion and calcium/blooming artifacts; ICA: inadequate view separation, overlap/foreshortening, and insufficient contrast filling), as well as end-to-end operating time, learning curves, and stratified performance in complex morphologies (bifurcation, diffuse/serial disease). For PINNs/PINOs that output curves or fields, assessment should extend beyond pointwise FFR$\leq$0.80 classification to include curve/shape agreement and decision-impact–relevant endpoints, to validate that they deliver interpretable, actionable physiology within cath-lab time constraints.

**Future directions.** Looking ahead, the priority is deployable system capability: stable training on branching geometries, systematic external validation across acquisition variability, and clinically aligned evaluation and reporting. In the near term, hybrid pipelines that combine physics-informed operators with reduced-order physics and vessel-tree representations are likely to be the most practical path to adoption.

## Funding

This research was supported by a research seed grant from Michigan Technological University Health Research Institute (PI: Weihua Zhou) and grants from the National Institutes of Health, USA (1R15HL172198 and 1R15HL173852) and American Heart Association (#25AIREA1377168).